\documentclass[fleqn,10pt]{wlscirep}
\usepackage[utf8]{inputenc}
\usepackage[T1]{fontenc}
\usepackage{lineno}
\usepackage{siunitx} %[binary-units]
\usepackage{hyperref}

\usepackage{multirow}       % multirow tables
\usepackage{bm}
\usepackage{amssymb}
\usepackage{onimage}            
\usepackage{tikz}

\PassOptionsToPackage{numbers, compress, super}{natbib}
\usepackage{natbib}
\usepackage{booktabs}       % professional-quality tables
\usepackage{pdflscape}
\usepackage{graphics}
\usepackage{wrapfig}
\usepackage{multirow}       % multirow tables
\usepackage{pifont}         % defining cmark and xmark
\newcommand{\cmark}{\ding{51}}%
\newcommand{\xmark}{\ding{55}}%
\usepackage{onimage}            
\usepackage{tikz} 
\usepackage{siunitx}
\usepackage{tabularx}
\DeclareMathOperator*{\argmin}{arg\,min}

\title{Benchmarking Learned Algorithms for Computed Tomography Image Reconstruction Tasks}

\author[1,$\dag$,*]{Maximilian B. Kiss}
\author[2,$\dag$]{Ander Biguri}
\author[2]{Zakhar Shumaylov}
\author[2]{Ferdia Sherry}
\author[1,3]{K. Joost Batenburg}
\author[2]{Carola-Bibiane Schönlieb}
\author[1]{Felix Lucka}
\affil[1]{Centrum Wiskunde \& Informatica, Computational Imaging group, Amsterdam, 1098 XG, The Netherlands}
\affil[2]{University of Cambridge, DAMTP, Cambridge, CB3 0WA, United Kingdom}
\affil[3]{Leiden University, LIACS, Leiden, 2300 RA, The Netherlands}

\affil[*]{corresponding author(s): Maximilian B. Kiss (maximilian.kiss@cwi.nl)}
\affil[$\dag$]{these authors contributed equally to this work}

\begin{abstract}
Computed tomography (CT) is a widely used non-invasive diagnostic method in various fields, and recent advances in deep learning have led to significant progress in CT image reconstruction. However, the lack of large-scale, open-access datasets has hindered the comparison of different types of learned methods. To address this gap, we use the 2DeteCT dataset, a real-world experimental computed tomography dataset, for benchmarking machine learning based CT image reconstruction algorithms. We categorize these methods into post-processing networks, learned/unrolled iterative methods, learned regularizer methods, and plug-and-play methods, and provide a pipeline for easy implementation and evaluation. Using key performance metrics, including SSIM and PSNR, our benchmarking results showcase the effectiveness of various algorithms on tasks such as full data reconstruction, limited-angle reconstruction, sparse-angle reconstruction, low-dose reconstruction, and beam-hardening corrected reconstruction. With this benchmarking study, we provide an evaluation of a range of algorithms representative for different categories of learned reconstruction methods on a recently published dataset of real-world experimental CT measurements. The reproducible setup of methods and CT image reconstruction tasks in an open-source toolbox enables straightforward addition and comparison of new methods later on. The toolbox also provides the option to load the 2DeteCT dataset differently for extensions to other problems and different CT reconstruction tasks.
\end{abstract}
\begin{document}

\flushbottom
\maketitle

\thispagestyle{empty}

\section*{Introduction}
Computed tomography (CT) is a widely used non-invasive diagnostic method which is applied in various fields such as medicine, materials science, industrial testing, and cultural heritage research. Based on X-ray projection images acquired from 360 degrees, cross-sectional images of an object or patient can be calculated using computer algorithms. Typically, a variety of image processing techniques are necessary to improve reconstruction quality of CT data that is, for example, badly sampled (limited/sparse-angle), low-dose, or exhibits artifacts caused by e.g. metals or other dense materials.
\\[6pt]
The rise of deep learning \cite{lecun2015deep} and widespread availability of large scale computing systems have led to substantial advances in computer vision, including tasks such as object detection, classification, segmentation or image denoising. A key enabler were the releases of corresponding large-scale, open-source datasets such as MNIST \cite{lecun1998gradient}, CIFAR \cite{krizhevsky2009learning} and ImageNet \cite{deng2009imagenet}, which helped the research community to hill-climb on standardized benchmarks and continuously advance the state-of-the-art. The last five years have also seen a rapid development of machine learning approaches specifically for CT image reconstruction \cite{adler2017solving,jin2017deep,chen2017low,wang2018image,wurfl2018deep,yang2018low,park2018ct,arridge2019solving,shan2019competitive,li2020nett,leuschner2021quantitative}, which hold great potential for further reducing patient dose, speeding up acquisitions, and improving image quality in challenging acquisition settings \cite{brady2021improving,gong2021deep}. 
\\[6pt]
But despite the increasing research activity at the intersection of CT and machine learning, the field of CT image reconstruction still lacks large-scale, open-access, real-world datasets that employ standardized evaluation metrics and benchmarking baselines. Many CT studies use datasets that are either not openly available to the research community or largely consist of synthetic data, which may suffer from the broadly observed sim-to-real gap \cite{nikolenko2021synthetic,weiss2024simulation}. Furthermore, most of them use different pre-processing pipelines and datasets of various sizes. This hinders the comparison of different state-of-the-art methods and makes reproducing as well as validating results a cumbersome and challenging task.
\\[6pt]
Early computer vision algorithms were often developed using small-scale datasets under lab conditions and showed a significant lack of generalization in the real world. Generalization ability improved with the advent of large-scale datasets composed of images from the internet, which then grew to include increasingly unstructured data, to today's massive multi-modal (visual and language) datasets scraped off the internet enabling unprecedented forms of information processing and synthesizing by foundational models and LLMs. Accordingly, developing a large-scale benchmarking dataset for 2D computed tomography may be a first step to enabling similar breakthroughs in data-driven CT image reconstruction.
\\[6pt]
In this paper, we utilize real-world experimental data instead of simulated CT data and design standardized experiments for various common CT reconstruction tasks, allowing for more systematic and standardized comparisons between learned algorithms on one unified dataset. The contributions of this paper are: (i) a benchmarking study for a fixed set of data-driven methods on a recently published dataset of real-world experimental measurements, the 2DeteCT dataset \cite{Kiss_23}; (ii) a toolbox for benchmarking that enables the seamless addition of new methods; (iii) an option to load the 2DeteCT dataset differently within the toolbox for extensions to other problems and different CT reconstruction tasks. It provides a starting point for the community to develop, test, and compare new methods on real-world experimental data in a straightforward and reproducible way, which can shorten the overall development time of new data-driven CT image reconstruction algorithms considerably. 
\\[6pt]
In the remainder of this paper, we give a brief overview of related work in the field of CT datasets, provide details about the mathematical foundation of data-driven CT reconstruction and dive into a quick categorization of learning-based methods for solving inverse problems such as CT image reconstruction to give context for the information content of the benchmarking framework. Afterwards, we introduce the benchmarking design, including the various CT image reconstruction tasks, the data pipeline, and the performance metrics. Subsequently, we elaborate on the employed pre-processing of the benchmarking dataset, the evaluated methods of our numerical experiments, and their training details. After presenting the benchmarking results we discuss the limitations of the dataset, its broader impact, and the code and data availability.

\begin{table}[!ht]
    \caption{A summary of publicly available CT datasets, supported tasks, their size, and their raw data availability. (\cmark) = possible through data generation.} % not natively supported
    \label{table:datasets-summary}
    \centering
    \resizebox{1.0\linewidth}{!}{
    \begin{tabular}{lcccccc}
        \toprule
        \multirow{2}{*}{\textbf{Dataset}} & \multicolumn{4}{c}{\textbf{CT Image Reconstruction Tasks}} & \multirow{2}{*}{\textbf{Size (\textgreater{}100 samples)}} & \multirow{2}{*}{\textbf{Raw Data}} \\
        \cline{2-5}
        & \textbf{Low-Dose} & \textbf{Limited-Angle} & \textbf{Sparse-Angle} & \textbf{Beam hardening reduction} \\
        \midrule
        Mayo \cite{mccollough2016tu,moen2021low} & \cmark & (\cmark) & (\cmark) & \xmark & \xmark \cite{mccollough2016tu} / \cmark \cite{moen2021low} & \xmark \\
        LoDoPaB \cite{leuschner2021lodopab} & \cmark & (\cmark) & (\cmark) & \xmark & \cmark & \xmark \\
        ICASSP GC8 \cite{biguri2024advancing} & \cmark & (\cmark) & (\cmark) & \xmark & \cmark & \xmark \\
        Walnut CBCT \cite{Sarkissian_2019} & \xmark & \cmark & \cmark & \xmark & \xmark & \cmark \\
        2DeteCT \cite{Kiss_23} & \cmark & \cmark & \cmark & \cmark & \cmark & \cmark \\
        \bottomrule
    \end{tabular}
    }
\end{table}

\section*{Related Work} \label{sec:related_Work}
Computer science has played a vital role in overcoming limitations of traditional imaging systems such as CT and magnetic resonance imaging (MRI). In combination with applied mathematics and advanced engineering, computer science forms the field of computational imaging. The main goal of computational imaging is to improve image quality, enhance resolution, enable novel imaging capabilities, and extract valuable information or hidden details and features that may not be directly visible by traditional imaging methods.
\\[6pt]
The field has undergone many technological advances throughout the last 15 years \cite{wang2008outlook,ritman2011current,ginat2014advances,hsieh2021computed} but the most recent focus has been on employing machine learning techniques \cite{wang2018image}. Despite its clear necessity, the computational imaging field to date offers few large-scale datasets and benchmarks on real-world experimental data. Researchers of NYU and Facebook recently addressed this need for the case of MRI scans by publishing raw measurement data in their fastMRI \cite{Knoll_2020} dataset, while the field of CT still lacked an open-access dataset of comparable scope (cf. Table \ref{table:datasets-summary}). Acquiring such data in the medical sector presents particular challenges, including the radiation exposure patients receive from multiple CT scans and the lack of access to raw measurement data from commercial CT scanners. Previous attempts in the field of low-dose CT, such as the Mayo Clinic low-dose CT challenge of 2016 \cite{mccollough2016tu} and 2021 \cite{moen2021low}, the LoDoPaB dataset \cite{leuschner2021lodopab}, and the the IEEE ICASSP Grand Challenge 8 \cite{biguri2024advancing} have sought to bridge this gap, but relied on simulated data. These issues are further exacerbated due to the lack of raw projection data along with corresponding reconstructed image slices. Although the second release of the Mayo Clinic low-dose CT challenge of 2021 \cite{moen2021low} already released raw projection data, the noise for the low-dose datasets remains simulated.
\\[6pt]
Only recently, the 2DeteCT dataset \cite{Kiss_23} overcame these shortcomings by providing raw measurement data with complementary features that can be used for a wide range of imaging tasks such as supervised or unsupervised denoising, limited- and sparse-angle scanning, beam-hardening reduction, super-resolution, region-of-interest tomography or segmentation. In contrast to the clinical, in-vivo dataset such as the LIDC-IDRI \cite{armato2011lung} or Mayo Clinic datasets \cite{mccollough2016tu,moen2021low}, the 2DeteCT dataset would be categorized as an in-vitro dataset that only simulates the behavior of natural tissue.
\\[6pt]
However, having one universal dataset instead of individual datasets of various research groups helps to train algorithms on a uniform set of data, to test them in a standardized way, and to compare them against other algorithms for different imaging tasks. Particularly, the problem of defining a ground truth or ``gold standard'' is prevalent in CT imaging. Usually, it involves some sort of choice or trade-off with respect to the image acquisition or generation whereas for the 2DeteCT dataset the ``mode 2'' acquisition provides clean data since its acquisition was designed in a high-resolution setting with an over-sampling in the number of angular projections, a high-dose tube setting, and with a beam filtration in place. Therefore, we treat the reference reconstructions of the 2DeteCT dataset utilizing a cropped Nesterov gradient descent (AGD) as a ground truth or ``gold standard'' in this work. They can be used as target images for the matching noisy or artifact-inflicted measurements of ``mode 1'' and ``mode 3'' respectively and for limited- or sparse-angle measurement data extracted from ``mode 2''. A more detailed description of these acquisition modes can be found in section ``Relevance and Difficulty of CT Image Reconstruction Tasks'' and a visualization is presented in Figure \ref{fig:CTReconstructionTasks}. This figure also illustrates the types of artifacts present in each reconstruction, highlighting the diverse challenges that data-driven CT reconstruction must address.

\section*{Data-driven CT Reconstruction}
In this section, we present the mathematical background of data-driven CT reconstruction, specifically focusing on 2D tomography, i.e. reconstructing 2D slices from 1D projection data. Furthermore, we briefly introduce how the four classes of methods explored in this work fit within this framework.

\subsection*{Tomographic Reconstruction as a Linear Inverse Problem}
Tomographic reconstruction is an inverse problem which can be described as an image recovery task based on measurements obtained through the Radon transform: $y(\ell)=\int_{\ell}x(z)\,\mathrm{d}z, \ell\in\mathcal{L}$. In this equation, $\mathcal{L}$ represents the lines in $\mathbb{R}^2$ from the X-ray source to each detector pixel, defined by the scanner geometry and rotation. Typically, this problem is linearized and discretized as
\begin{equation}
    Ax + \tilde{e} = y  \label{eq:linear}
\end{equation}
where $A$ represents the so-called \textit{forward operator} which encapsulates the integral computations over these lines. Here, $A$ is a matrix where each row corresponds to a line integral over the pixel grid of the object. In this context, $x$ is a vector representing the pixel values of the image, $y$ is a vector representing the measured sinogram values, and $\tilde{e}$ accounts for the noise or error, which may arise from the measurements themselves or from the linearization of the operator.
\\[6pt]
Classically, to solve the inverse problem in Eq.~\ref{eq:linear} in a robust manner, a variational regularization approach \cite{englRegularizationInverseProblems2000, scherzer2009variational} is employed. The reconstruction is defined by the following minimization problem of the variational objective:
\begin{equation}
\hat{x} = \argmin_x \left\{ \mathcal{D}(y,Ax) + \mathcal{R}(x)\right\},\label{eq:vr}   
\end{equation}
where $\mathcal{D}$ measures the data fidelity between the measurement and the reconstructed image (most commonly the $L^2$-distance in CT) and $\mathcal{R}$ is a regularization function that promotes images of desired properties. The data fidelity term $\mathcal{D}$ is usually chosen according to the noise distribution, and a good choice of regularizer $\mathcal{R}$ is important for achieving accurate results. Traditionally, regularization functionals were hand-crafted to encourage the reconstruction $x$ to have structures known to be realistic.
\\[6pt]
In practice, Eq.~\ref{eq:vr} is solved using iterative optimization schemes, and the quality of reconstructions is largely influenced by the choice of the regularization functional $\mathcal{R}$. A variety of methods have been proposed in the optimization literature to solve Eq.~\ref{eq:vr} with particular choices for the data fidelity term $\mathcal{D}$ and the regularization term $\mathcal{R}$, often under the assumption that these functions are convex. In certain cases, these optimization methods yield superior reconstructions compared to the standard analytical approach of the inverse Radon transform, known as filtered backprojection, especially when suitable functions and parameters are selected.
\\[6pt]
While convexity of $\mathcal{R}$ is analytically desirable for providing efficient optimization schemes with various guarantees, it is often observed that non-convex regularizers yield superior reconstructions in practice. However, this advantage comes at a price: finding global minima becomes generally infeasible, and sometimes even finding stationary points cannot be guaranteed.

\subsection*{Data-driven Methods for Tomographic Reconstruction}
\label{sec:data-driven_methods}
In response to the limitations of classical knowledge-driven approaches, data-driven methods have rapidly advanced over the past decades. These methods can be categorized in different ways, including based on the amount of expert knowledge involved, which components are parameterized as neural networks, the domain of application, and the methodological approach employed (cf. Table \ref{table:method_categorizations}). In this work, we follow the general categorization of supervised learning methods of Arridge et al. \cite{arridge2019solving} and consider the following methods in our benchmarking design: post-processing networks, learned/unrolled iterative methods, learned regularizer methods, plug-and-play methods.

\begin{table}[!ht]
  \caption{Method Categorizations}
  \label{table:method_categorizations}
  \centering
  \vspace{0.2cm}
  \resizebox{0.95\linewidth}{!}{
  \begin{tabularx}{\textwidth}{lX}
    \toprule
    Article     & Categorization\\
    \midrule
    Zhang et al., 2020 \cite{zhang2020review} & Amount of expert knowledge involved: handcrafted, hybrid approaches, mostly learned \\
    Ye et al., 2023 \cite{ye2023deep} & Point of learned processing: pre-processing, post-processing, and raw-to-image \\
    Ravishankar et al., 2019 \cite{ravishankar2019image} & Domain of application: image-domain, hybrid-domain, AUTOMAP \cite{zhu2018image}, sensor-domain \\
    Arridge et al., 2019 \cite{arridge2019solving} & Methodological: Post-processing Networks, Learned / Unrolled Iterative Methods, Learned Regularizer Methods, Plug-and-Play Methods \\
    \bottomrule
  \end{tabularx}
}
\end{table}
\noindent
While any strict categorization may overlook or misrepresent certain methodologies from the literature, such as self-supervised learning strategies, this framework effectively encompasses the majority of techniques found in the (weakly) supervised learning literature on data-driven CT reconstructions.
\\[6pt]
In the following section, we briefly introduce each of the method categories and give a foundational description of them for historical and methodological context. It is not intended as a state-of-the-art review of methods in the literature.
\\[6pt]
\underline{Direct Solvers:} Directly learning a reconstruction from measurement $y$ as $\hat{x}=\mathcal{N}_\theta(y)$ has been proposed for MRI in AUTOMAP~\cite{zhu2018image}. Recently, a similar direct method has been proposed for CT \cite{fu2022deep}, but due to the limited success of this approach, we do not explore it further.
\\[6pt]
\underline{Post-processing Networks:} A straightforward approach to incorporating data-driven methods to overcome the ill-posed nature of CT reconstruction is to initially use classical method for reconstruction and subsequently train a network to learn the mapping from the manifold of inadequate reconstructions to the manifold ground truth reconstructions\cite{kang2017deep,jin2017deep,han2018framing,zhang2018sparse,gupta2018cnn}:
\begin{equation}
    \hat{x}= \mathcal{N}_\theta(\mathcal{F}(y)),
\end{equation}
where $\mathcal{F}$ is a reconstruction (e.g. FBP in this work), and $\mathcal{N}_\theta$ an appropriately parameterized neural network (NN).
\\[6pt]
\underline{Learned/Unrolled Iterative Methods:} This method category arises from the realization that many iterative solvers, such as FISTA~\cite{beck2009fast}, have a resemblance to convolutional neural networks (CNNs)~\cite{gregor2010learning}. Notably, Barbu~\cite{barbu2009training} introduced the idea of unrolling iterative methods. By unrolling a handcrafted iterative algorithm and using it as a building block of a deep neural network (DNN), the parameterized mathematical operators inherit hyperparameters, image priors, and data consistency constraints from the iterative methods \cite{ye2023deep}. A common way to design these methods is to find a particular step of an iterative solver, e.g. a \textit{proximal} step, and replace it with an iteration-dependent shallow CNN. A broader overview of these methods is given in the review by Monga et al.~ \cite{monga2021algorithm}.
\\[6pt]
The resulting unrolled iterative methods have more interpretable architectures compared to traditional ``black-box'' denoisers, commonly exhibit much fewer trainable parameters than standard DNNs, and enable combining domain knowledge with deep learning \cite{zhang2020review}. However, it is important to acknowledge that while these methods may be more interpretable, they lose their mathematical guarantees when incorporating learned networks. Since these methods utilize the operator within the network, they are often referred to as model-based networks. In this context, the physics of the model, represented by the operator $A$, is provided to the network rather than learned.
\\[6pt]
\underline{Learned Regularizer Methods:} Fundamentally, this approach is based on learning the regularization functional $\mathcal{R}$ in Eq. \ref{eq:vr} from data. Traditionally, the regularization functional was hand-crafted for the problem, and over the past decades many hand-crafted functionals have been proposed; see Benning and Burger \cite{benning2018modern} for an overview. In contrast to standard DNNs, the minimization of the variational objective can be analyzed mathematically \cite{mukherjee2023learned}. Examples include dictionary learning \cite{chen2016compressed}, deep image priors \cite{DIP}, generative \cite{webber2024diffusion}, network Tikhonov \cite{li2020nett}, and adversarial regularization \cite{shumaylov2024weakly}. See \cite{habring2024neural,dimakis2022deep} for an overview.
\\[6pt]
\underline{Plug-and-Play Methods:} A special sub-case of learned regularizers is the field of Plug-and-Play (PnP) methods where proximal algorithms are used to optimize the inverse problem when either the data fidelity or regularization term is non-smooth. Two widely used iterative algorithms minimizing such composite functionals are the Alternating Direction Method of Multipliers (ADMM) \cite{boyd2011distributed} and the FISTA \cite{beck2009fast} which use proximal operators to avoid differentiating the non-smooth function. The proximal step in these algorithms can be replaced by a more general black-box denoiser (``plugged''-in) while the optimization algorithms run (``play'') as before. This approach of PnP methods was developed by Venkatakrishnan et al. \cite{venkatakrishnan2013plug} and an overview of theory, algorithms, and applications can be found in a recent review by Kamilov et al. \cite{kamilov2023plug}. Although PnP methods are heavily inspired by variational approaches, the study of their properties as convergent regularization methods is an area that is still under active development, with some initial work establishing results in this direction \cite{ebner2024plug,hauptmann2023convergent}.

\section*{Benchmark Design} \label{sec:benchmark_design}
\begin{figure}[htp]
  \centering
  \includegraphics[width=1\linewidth]{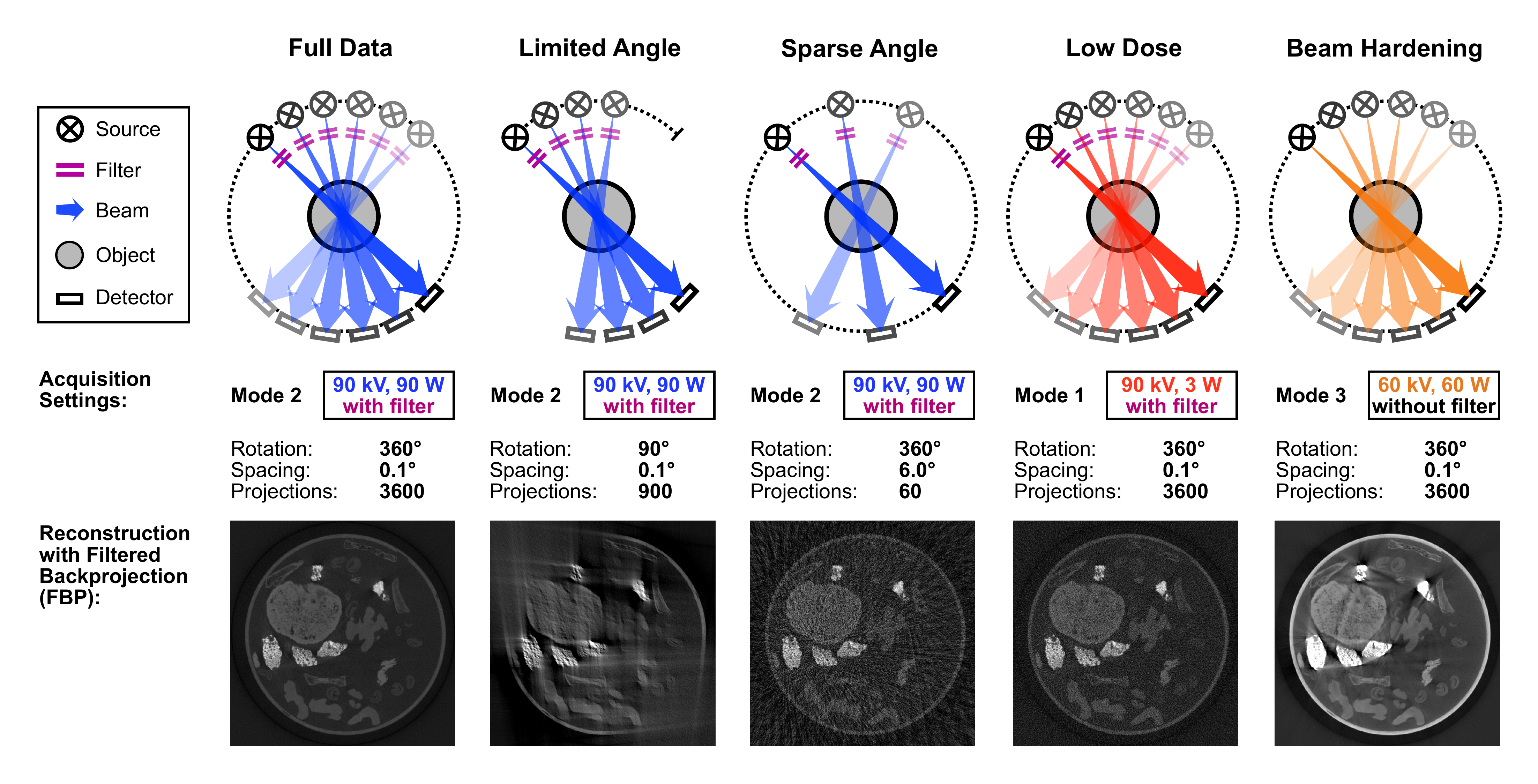}
  \caption{CT Image Reconstruction Tasks.}
  \label{fig:CTReconstructionTasks}
\end{figure}

Following best practices on reproducibility for benchmarks \cite{weber2019essential} we define the purpose and scope of our benchmark as providing the research community with a benchmarking framework based on a real-world experimental dataset under CC BY 4.0 license. It consists of a versatile toolbox as well as a pipeline to evaluate and compare different algorithms and setting up reproducible and reusable experiments for different image reconstruction and processing tasks in X-ray computed tomography. The methods selected cover the full range of common categories of supervised learning methods for solving inverse problems. For each category we implement three well-established methods and evaluate their respective performance. The parameters for the CT image reconstruction tasks represent common choices in the field and all methods are implemented in recent Python and PyTorch versions. The evaluation is done with key quantitative performance metrics such as the structural similarity index (SSIM) and peak signal-to-noise ratio (PSNR). The results of the different methods are documented in Tables \ref{table:quantitative_analysis_modes}, \ref{table:quantitative_analysis_limited}, and \ref{table:quantitative_analysis_sparse} as well as in a visual overview for a selected image slice in Figure \ref{fig:qualitative_analysis}. All trained models are saved and made available on GitHub. The LION toolbox \href{https://github.com/CambridgeCIA/LION/}{https://github.com/CambridgeCIA/LION/} used for setting up the benchmarking experiments enables future extensions by implementing other CT experiments or other ML-based methods within the LION framework. The codebase is open source and licensed under GNU General Public License v3.0. The aim of the benchmarking design is threefold: Firstly, we give an overview of the different categories of data-driven methods for CT image reconstruction. Secondly, we set-up an easy-to-use pipeline for implementing and testing algorithms on real-world experimental data. Thirdly, we provide a baseline comparison of the aforementioned data-driven methods on the most common CT image reconstruction tasks \cite{ye2023deep}. 
\\[6pt]
The benchmark design described in this work is unique in its combination of realism, dataset scale, variety of measurement settings and variety of reconstruction methods considered. Our major contribution is that we have developed a benchmarking framework that for the first time relies completely on real-world experimental data and investigates the whole range of common CT image reconstruction tasks. Existing studies on data-driven CT reconstruction usually focus on one of the method categories and a singular task for which a newly developed algorithm is compared to the most recent state-of-the-art and classical reconstruction methods. Often the data used for these assessments are not the same as the data the other method was tested on, e.g. the acquisition geometry, the sub-sampling, or the pre-processing might differ. All these factors limit their comparability and necessitate a benchmarking design with one common dataset and standardized CT reconstruction tasks as outlined below and visualized in Figure \ref{fig:CTReconstructionTasks}.

%\begin{wrapfigure}{R}{0.5\linewidth}
%\end{wrapfigure}

\subsection*{CT Image Reconstruction Tasks}
2DeteCT encompasses raw experimental data of various acquisition modes for different CT image reconstruction tasks unified in one dataset. Defining and setting up CT image reconstruction tasks from the 2DeteCT dataset requires loading subsets of the experimental measurement data and defining the geometry for the reconstructions (see Table \ref{tab:acquisition_parameters}). For this work, we use the 2DeteCT sinograms of shape $956\times3600$ and reference reconstructions of shape $1024\times1024$.

\begin{table}[ht]
\centering
\caption{\label{tab:acquisition_parameters} Summary of the acquisition parameters of the 2DeteCT dataset, adapted from \cite{Kiss_23}. *$^1$ (Thoraeus = Sn 0.1mm, Cu 0.2mm, Al 0.5mm), *$^2$ these quantities are based on the motor readings of the FleX-ray scanner which get translated into physical quantities and are subject to alignment errors.}
\resizebox{0.55\linewidth}{!}{
\begin{tabular}{|c|c|c|c|}
\hline
\textbf{Acquisition parameter} & \textbf{Mode 1} & \textbf{Mode 2} & \textbf{Mode 3} \\
\hline
Tube voltage & $\SI{90.0}{\kilo\volt}$ & $\SI{90.0}{\kilo\volt}$ & $\SI{60.0}{\kilo\volt}$  \\
\hline
Tube power & $\SI{3.0}{\watt}$ & $\SI{90.0}{\watt}$ & $\SI{60.0}{\watt}$  \\
\hline
Filters used & Thoraeus*$^1$ & Thoraeus & No Filter \\
\hline
Exposure time & \multicolumn{3}{c|}{$\SI{50.0}{\milli\second}$} \\
\hline
Binned detector pixel size & \multicolumn{3}{c|}{$\SI{149.6}{\micro\metre}$} \\
\hline
Number of binned detector pixels & \multicolumn{3}{c|}{956} \\
\hline
Source to object distance *$^2$ & \multicolumn{3}{c|}{$\SI{431.020}{\milli\metre}$} \\
\hline
Source to detector distance *$^2$ & \multicolumn{3}{c|}{$\SI{529.000}{\milli\metre}$} \\
\hline
Number of projections & \multicolumn{3}{c|}{3601} \\
\hline
Angular increment & \multicolumn{3}{c|}{$\SI{0.1}{\deg}$} \\
\hline
\end{tabular}
}
\end{table}

\noindent
Each of the tasks utilizes different properties of the 2DeteCT dataset, necessitates a corresponding data pairing and requires a pre-processing which we will discuss in Section ``Prep-processing''. The ``gold standards'' set as target images for all tasks are the reference reconstructions of ``mode 2'' of the 2DeteCT dataset which utilize a cropped Nesterov gradient descent (AGD). The details of these tasks are described in the following and visualized in Figure \ref{fig:CTReconstructionTasks}.
\\[6pt]
\underline{Full Data Reconstruction:} We use the complete raw projection data of the ``mode 2'' acquisition (full sinograms) and use the corresponding iterative reference reconstructions of the 2DeteCT dataset as target images. Since these were generated through a Nesterov Accelerated Gradient (AGD) Descent algorithm on a bigger reconstruction plane and cropped to its center, the evaluated methods actually learn to mimic AGD for this task. However, the Full Data Reconstruction serves as a reference for the performance of the evaluated methods on other CT image reconstruction tasks.
\\[6pt]
\underline{Limited-Angle Reconstruction:} We limit the raw projection data of the ``mode 2'' acquisition to a smaller angular range. Depending on the used wedge of $120^\circ$, $90^\circ$, or $60^\circ$ only the first 1200, 900 or 600 projection lines of the $956\times3600$ sinograms are extracted. With this missing information a standard reconstruction will show streaking, elongation, ghost tail, and missing boundaries artifacts of increasing severity with decreasing angular size of the wedge \cite{frikel2013characterization,barutcu2021limited}. The ``gold standard'' is a reconstruction of a complete sinogram and therefore we use the iterative reference reconstructions of the 2DeteCT dataset as target images.
\\[6pt]
\underline{Sparse-Angle Reconstruction:} We sub-sample the raw projection data of the ``mode 2'' acquisition. Depending on the number of angles used $360$, $120$, or $60$ projection lines evenly distributed over the full angular range are extracted from the original $3600$ projections of the sinograms. Undersampling of this kind violates the Nyquist-Shannon sampling theorem and introduces aliasing artifacts \cite{joseph1980view}. The ``gold standard'' is a reconstruction of a fully-sampled sinogram and therefore we use the iterative reference reconstructions of the 2DeteCT dataset as target images.
\\[6pt]
\underline{Low-Dose Reconstruction:} We use the complete raw projection data of the ``mode 1'' acquisition (full sinograms). Due to the low-dose setting of the acquisition the measurements show very low photon counts and their corresponding iterative reconstruction slices show streaking artifacts and high granularity as manifestations of the image noise. The ``gold standard'' for this task is a reconstruction of the corresponding high-dose acquisition of ``mode 2'' and we use the iterative reference reconstructions of the 2DeteCT dataset as target images.
\\[6pt]
\underline{Beam-Hardening corrected Reconstruction:} We use the complete raw projection data of the ``mode 3'' acquisition (full sinograms). Due to the unfiltered beam spectrum of the X-ray source during the acquisition of the measurements the corresponding iterative reconstruction slices show streaking and shadowing-like ``cupping'' artifacts as manifestations of beam hardening and photon starvation \cite{patton2008spect,van2002energy}. The ``gold standard'' for this task is a reconstruction of the corresponding filtered acquisition of ``mode 2'' and we use the iterative reference reconstructions of the 2DeteCT dataset as target images.

\subsection*{Pipeline} \label{sec:pipeline}
The benchmark framework was set up using LION (Learned Iterative Optimization Networks), an open-source Python toolbox for learned tomographic reconstruction. There are other open-source libraries available that focus on providing a robust set of forward (and backward) operators, like ODL \cite{jonas_adler_2017_556409} or tomosipo \cite{hendriksen2021tomosipo} or libraries that focus on providing variational (non-ML) reconstructions, such as ASTRA \cite{PALENSTIJN2011250,VANAARLE201535,vanAarle16} or TIGRE \cite{biguri2016tigre,biguri2020arbitrarily}. Given such existing libraries, obtaining an operator for the experiments described above, can be straightforward, but creating such operators for real measured datasets is not always trivial and loading this data accordingly can be challenging. LION focuses on using libraries such as tomosipo to build a toolbox for loading, pre-processing, simulating, and reconstructing CT data as well as training and evaluating data-driven methods in CT. As forward operators, LION currently supports the aforementioned tomosipo, which uses ASTRA's ray-driven forward projector and an unmatched voxel-driven backprojector.
\\[6pt]
To our knowledge there is only one other maintained open-source library for deep learning that supports CT reconstruction, DeepInv \cite{tachella2023deepinverse}. DeepInv is a comprehensive library for inverse problems and learning. However, it does have limitations in terms of specificity, which is often overlooked by more generalist libraries. For instance, it does not provide complex dataset definitions, topic-specific experiments, or application specific noise models that are crucial for certain applications.
 \\[6pt]
Through this benchmarking study LION now features a designated data loader for the 2DeteCT dataset \cite{Kiss_23} and defines tasks and modes such that the data loader sources the corresponding data from this dataset. It performs a split into training (79.3\%, 3930 slices), validation (11.11\%, 550 slices), and test data (9.49\%, 470 slices), defines the CT geometry and parameters of CT data processing, creates a forward operator, and loads and pre-processes sinograms and reconstructions. Furthermore, it contains designated experiment classes for the above CT image reconstruction tasks and PyTorch implementations of various models from the different method categories outlined above. A significant development effort was undertaken to ensure a consistent implementation of all deep learning methods using the same data and operator framework for the variety of CT image reconstruction experiments. Additionally, the toolbox defines metrics for training and evaluation, contains an optimizer for the supervised learning setting and allows for saving all relevant information in a parameter file to completely reproduce models. These parameter files store among others, the used dataset parameters, training parameters, loss, epochs, optimizer, and the CT geometry. Lastly, the toolbox allows for saving trained models to compare against them and for storing scripts that have been used in papers to specifically reproduce the experiments of that particular study.
\\[6pt]
To summarize, the motivation for this benchmarking design is providing the community with an easy pipeline to load this data and conduct standardized experiments is the unique contribution of this benchmarking study. It lays the foundation for the computational imaging community to easily implement and test methods on the 2DeteCT dataset using a specific data loader as well as tailored and standardized CT benchmarking experiments. This greatly extends the utility of the 2DeteCT dataset since researchers do not need to spend time on the implementation of their own data loaders or reconstruction tasks and can easily compare against other methods. It makes it easier than ever to start experimenting with deep-learning-based (and non-deep-learning-based) CT reconstruction in realistic settings, without the need for expert knowledge or simulating data. In particular, it allows users to avoid many of the pitfalls of trying to simulate appropriate measurement data and focus instead on the development of reconstruction methods.

\subsection*{Performance Metrics}
One common metric for evaluating CT reconstructions especially in the case of limited or noisy data is the peak signal-to-noise ratio (PSNR)\cite{girod1992psychovisual}. It quantifies the ratio of the maximum possible value of a signal to the power of corrupting noise that affects the fidelity of the image. Furthermore, the structural similarity (SSIM) \cite{wang2004image,sheikh2006statistical,venkataramanan2021hitchhiker} indicates in a range from 0.0 to 1.0 how similar the evaluated image is to this reference image, where 1.0 means they are identical. For both metrics, higher scores indicate a better algorithm performance and a ground truth reference image is necessary. The ground truth reference image used in this work are the reference reconstructions of ``mode 2'' of the 2DeteCT dataset which utilize a cropped Nesterov gradient descent (AGD).

\section*{Numerical Experiments} \label{sec:numerical_experiments}

\subsection*{Pre-processing} \label{sec:pre-proc}
All numerical experiments for the different CT image reconstruction tasks are set up as sinogram-to-reconstruction ``sino2recon'' experiments. They do not perform sinogram-to-sinogram ``sino2sino'' or reconstruction-to-reconstruction ``recon2recon'' experiments such as sinogram denoising, artifact reduction, inpainting, beam hardening reduction. This means that the ML-based algorithms take a sinogram as input data and corresponding iterative reference reconstructions from acquisition ``mode 2'' as target data. In principle, it is possible to also perform ``sino2sino'' or ``recon2recon'' experiments within the LION toolbox but this would make the comparison between e.g. post-processing networks and learned/unrolled iterative methods less fair. Therefore, we chose to only use ``sino2recon'' experiments in this benchmarking study. The sinograms are pre-processed with LION using modules such as \href{https://astra-toolbox.com}{ASTRA} \cite{PALENSTIJN2011250,VANAARLE201535,vanAarle16} and tomosipo \cite{hendriksen2021tomosipo} according to the description in the original dataset publication and as outlined below.
\\[6pt]
The sinograms are pre-processed into a beam intensity loss image by subtracting detector off-set counts (“dark currents”) from the measured photon counts per detector pixel and by dividing by the so-called ``flat fields'', the pixel-dependent sensitivities of the detector. To perform a CT reconstruction the data then is transformed with the negative logarithm to follow the Beer-Lambert law. For more details please refer to the original dataset publication \cite{Kiss_23}.

\subsection*{Evaluated Methods}
\label{sec:evaluated_methods}
The method selection in this work focuses on (weakly) supervised learning methods, excluding self-supervised and unsupervised approaches to establish a foundation for benchmarking. To this end, we prioritize established supervised learning methods that can serve as reliable baselines, omitting some newer techniques based on transformers and generative models, as explained at the end of this subsection. In the following, we introduce which methods from the literature will be used in this benchmark. To limit the broad scope of this work, we use three methods from each subclass presented in Section ``Data-driven Methods for Tomographic Reconstruction''.

%\begin{wraptable}{R}{.65\linewidth}
%\vspace{-0.68cm}
\begin{table}[!ht]
  \caption{Evaluated methods}
  \label{table:evaluated_methods}
  \centering
  \vspace{0.2cm}
  \resizebox{0.65\linewidth}{!}{
  \begin{tabular}{ll}
    \toprule
    Category     & Method (Year and Reference)\\
    \midrule
    Classical Methods    & FBP \cite{hansen2021computed}, AGD \cite{nesterov1983method}, ChP \cite{chambolle2011first} \\
    Post-Processing Methods         & U-Net \cite{ronneberger2015u}, MSD-Net \cite{pelt2018mixed}, DnCNN \cite{zhang2017beyond}      \\
    Learned / Unrolled Iterative Methods     & Learned Gradient \cite{adler2017solving}, TV-regularized Learned Gradient, \\ &  Learned Primal Dual \cite{adler2018learned} \\
    Learned Regularizer Methods     &  AR \cite{lunz2018adversarial}, TDV \cite{kobler2020total}, ACR  \cite{mukherjee2020learned,mukherjee2024data} \\
    Plug-and-Play Methods     & DnCNN-PnP \cite{zhang2017beyond}, DRUNet-PnP \cite{zhang2022plug}, GS-PnP \cite{huraultgradient} \\
    \bottomrule
  \end{tabular}
  }
\end{table}
%\vspace{-0.5em}
%\end{wraptable}
\noindent Table \ref{table:evaluated_methods} lists the different methods that have been evaluated for the benchmarking and in which of the described method categories they fall. The details of the networks' training can be found in Section `Training Details'' and in the GitHub repository mentioned in the section ``Code and Data Availability''. Additionally, we evaluate classical reconstruction methods on the test data. Analytical methods such as filtered backprojection (FBP) or iterative methods such as Nesterov accelerated gradient descent (AGD) \cite{nesterov1983method} or  regularized methods such as the Chambolle-Pock (ChP) \cite{chambolle2011first} solver with total variation (TV) regularization are highly effective and still widely used in practice \cite{beister2012iterative}.
\\[6pt]
\underline{Post-processing Networks:} In this work, the evaluated post-processing networks are the U-Net \cite{ronneberger2015u}, the MSD-Net \cite{pelt2018mixed}, and the DnCNN \cite{zhang2017beyond}. The U-Net, originally designed for image segmentation tasks, is well known across fields and consists of a contracting path, which captures both low and high-frequency features, a bottleneck layer, and a symmetric expanding path. The expanding path integrates information from corresponding layers in the contracting path, effectively translating the learned features back into the image space at each resolution.. The MSD-Net has proven itself particularly effective for CT image reconstruction problems in the literature. Its neural network architecture, incorporating dense connections between layers at different scales, helps to effectively capture both local and global information in images. The DnCNN is a deep learning architecture specifically designed for image denoising tasks. It uses a series of convolutional layers with batch normalization to learn noise patterns and remove them from images. It has achieved state-of-the-art results in image denoising benchmarks. 
\\[6pt]
\underline{Learned / Unrolled Iterative Methods:} The first unrolled method we consider is Learned Gradient (LG) \cite{adler2017solving}, which seeks to directly learn the update step in the gradient descent solver rather than relying solely on an additive step. The LG method parameterizes a fixed number of gradient steps to approximate the optimal direction based on the true gradient of the variational objective. The second method is an extension of this model including a TV regularization in its update rule (LGTV). In both of these methods, a small four-layer CNN is employed to replace the update step, taking the current image estimate and gradient(s) as inputs. The last method considered is the Learned Primal Dual (LPD) algorithm \cite{adler2018learned}. LPD solves the variational optimization problem by learning gradient steps in both primal and dual variables simultaneously. By jointly training two networks to update primal and dual variables, this algorithm can efficiently solve tasks such as image reconstruction and denoising and is known to produce high-quality results, using a minimal set of learned parameters. In LPD, each gradient step is substituted by a shallow four-layer CNN.
\\[6pt]
\underline{Learned Regularizer Methods:} Learned regularization methods, which directly parameterize the regularization functional using a neural network, typically differ in either their training strategy, network architecture, or variational objective optimization scheme. For evaluation, the adversarial regularizer (AR)\cite{lunz2018adversarial} and its convex counterpart (ACR)\cite{mukherjee2024data,mukherjee2020learned} are considered alongside total deep variation (TDV)\cite{kobler2020total}. Both adversarial regularizers are trained using a Wasserstein-1 distance-based loss, while TDV is trained by minimizing the distance between the ground truth and the reconstruction achieved via a fixed number of gradient steps on the variational objective. AR is parameterized using a standard CNN with a single dense layer, and the variational objective is optimized via accelerated gradient descent with early stopping. ACR utilizes an input convex neural network, and the variational objective is optimized with accelerated gradient descent and backtracking. TDV is parameterized using a multiscale convolutional neural network.
\\[6pt]
\underline{Plug-and-Play Methods:} There are various axes along which the settings of PnP methods can be varied, including the choice of splitting method and the architecture of the denoiser. In this work, we will fix the splitting to be a forward-backward splitting of a variational objective, and consider the effect of varying the denoiser architectures: two of them will be ``unconstrained'', differing mainly in model capacity, while the last one has a structural constraint that allows for provable convergence. To be more specific, the first method (DnCNN-PnP) replaces the proximal operator of the regularization functional by DnCNN~\cite{zhang2017beyond}, while the second method (DRUNet-PnP) uses a DRUNet \cite{zhang2022plug} instead, which gives improved denoising performance at the cost of significantly more parameters and increased computational time. Finally, the third method (GS-PnP) splits the variational objective in the opposite way, taking a gradient step on the regularization functional and a proximal step on the data discrepancy functional. It has been shown that it is possible to obtain high-quality PnP reconstructions in this way, while retaining the interpretation of minimizing a variational objective \cite{huraultgradient}. To compute the output of the denoiser in GS-PnP, it is necessary to perform an intermediate backpropagation on the backbone denoiser, resulting in significant extra computational cost, both in terms of memory and time. As in the work of Hurault et al. \cite{huraultgradient}, we deal with this by scaling down (both in number of blocks and width of the blocks) the backbone DRUNet, as compared to the DRUNet used in DRUNet-PnP.
\\[6pt]
The field of deep learning methods is rapidly evolving, with new architectures and methods constantly being released. For this reason, we necessarily have had to omit some methods from consideration in this benchmark such as the most recent strides using transformers and diffusion models for CT reconstruction. 
In the context of CT reconstruction, transformer-based reconstructions \cite{wang2023ctformer} generally take the form of what we have called a post-processing method in the benchmark, the only difference being the architecture of the “denoiser” used. In our benchmark, all of the architectures considered were convolutional. On the other hand, diffusion-model-based approaches \cite{song2022solving,liu2023dolce} are most similar to what we have called plug-and-play methods, as they alternately implement data-consistency steps, utilizing the forward model, and prior-consistency steps, which involve applying a denoiser in deterministic plug-and-play or executing a sampling step in stochastic restoration based on diffusion models.
\\[6pt]
In summary, we believe that our selection of method classes encompasses many methods of interest, even if some specific methods mentioned are not included in the presented comparison of twelve exemplary well-established approaches.

\subsection*{Training Details}\label{sec:modeltraining}
The basis for this benchmarking framework is the 2DeteCT dataset \cite{Kiss_23}. This dataset was split in a sophisticated way to ensure that no scanned sample mixes are shared between the training, validation, and test data. The data split is as follows: training data (79.4\%, 3,930 slices), validation data (11.11\%, 550 slices), test data (9.49\%, 470 slices). The training was carried out without extensive hyperparamter tuning to achieve an as good result as possible. We prioritized adequate performance over extensive hyperparameter tuning to produce baseline results for a comparative analysis among techniques. 

\subsubsection*{Post-Processing Methods}
\label{sec:postprocessing}
The post-processing networks have all been trained with the same parameters: Adam optimizer \cite{kingma2014adam} for 100 epochs with a learning rate of $10^{-4}$ and parameters $\beta_1=0.9$ and $\beta_2=0.99$. The final models were chosen based on the minimum loss in the validation set. The total training times are dependent on the used machine, the CT image reconstruction task and the evaluated method:
\begin{itemize}
    \item FBP+U-Net, total training time per CT image reconstruction task ranges \\ between $\sim 26-73$ hours,
    \item FBP+MSD-Net, total training time per CT image reconstruction task ranges \\ between $\sim 83-120$ hours,
    \item FBP+DnCNN, total training time per CT image reconstruction task ranges \\ between $\sim 56-93$ hours,
\end{itemize}

\subsubsection*{Learned / Unrolled Iterative Methods}
These methods have been trained exactly the same way as the post-processing methods, however, LG and LGTV required a learning rate of $10^{-5}$ for stable training. The final models were again chosen based on the minimum loss in the validation set. The total training times are dependent on the used machine, the CT image reconstruction task and the evaluated method:
\begin{itemize}
    \item LG, total training time per CT image reconstruction task ranges \\ between $\sim 24-117$ hours,
    \item LGTV, total training time per CT image reconstruction task ranges \\ between $\sim 19-116$ hours,
    \item LPD, total training time per CT image reconstruction task ranges \\ between $\sim 23-153$ hours,
\end{itemize}

\subsubsection*{Learned Regularizer Methods}
All models were trained using an Adam optimizer with a learning rate of $10^{-4}$ and parameters $\beta_1=0.9$ and $\beta_2=0.99$. These hyperparameters were picked according to the original methods \cite{mukherjee2020learned,kobler2020total}, and were fixed for all tasks. Order of magnitude for the hyperparameters in minimization of the variational objective were found for one tasks, the Sparse Angle Reconstruction with 90 projections, and were adjusted for other tasks based on the operator norm. The adversarial regularization methods have been trained for 25 epochs, with a reduced validation set. Minimum validation loss model was then chosen. TDV was trained for 10 epochs, due to its computationally expensive training. The number of steps was chosen to be the maximal number allowing for the network to fit on a 24 GB GPU. Due to the relatively low SSIM numbers we hypothesize that the number of steps ideally would need to be increased for all experiments, but due to the sizes, remains infeasible. The total training times are dependent on the used machine, the CT image reconstruction task and the evaluated method:
\begin{itemize}
    \item AR, total training time per CT image reconstruction task ranges \\ between $\sim 50-75$ hours,
    \item ACR, total training time per CT image reconstruction task ranges \\ between $\sim 50-75$ hours,
    \item TDV, total training time per CT image reconstruction task ranges \\ between $\sim 90-110$ hours,
\end{itemize}

\subsubsection*{Plug-and-Play Methods}
\label{sec:pnp_train}
As above, we trained the denoisers for the PnP methods using the Adam optimizer, with a learning rate of $10^{-4}$ and parameters $\beta_1=0.9$ and $\beta_2 = 0.999$. We normalized the inputs by rescaling with the maximum pixel value found on the training set. We trained denoisers on Gaussian denoising tasks with a range of noise levels $\{0.001, 0.005, 0.01, 0.02, 0.03, 0.05, 0.07\}$, corresponding to (average) PSNRs of approximately 50 dB, 36 dB, 30 dB, 24 dB, 20 dB, 17 dB and 13 dB respectively. After training, we plugged the denoisers into the methods described in Section ``Evaluated Methods'', selecting separately for each experiment the denoiser that performed best on the PnP reconstruction task on the validation set. Each denoiser was trained for 25 epochs, with total training times per denoiser being as follows:
\begin{itemize}
    \item DnCNN, total training time $\sim 6$ hours,
    \item DRUNet, total training time $\sim 10$ hours,
    \item GS-DRUNet, total training time $\sim 15$ hours.
\end{itemize}

\section*{Results and Discussion}
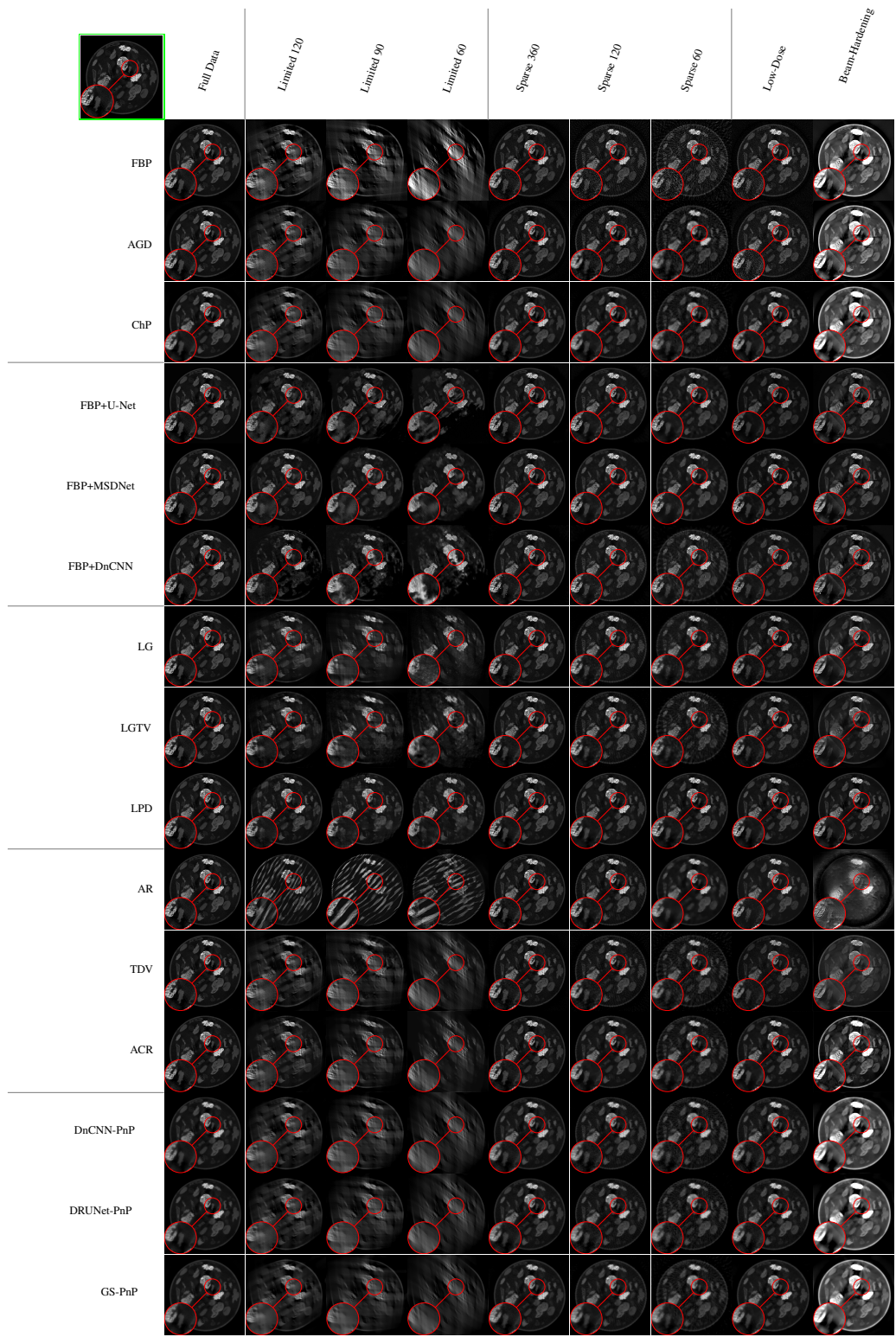
\begin{figure}[htp]
\centering

\begin{tikzonimage}[width=0.65\linewidth]{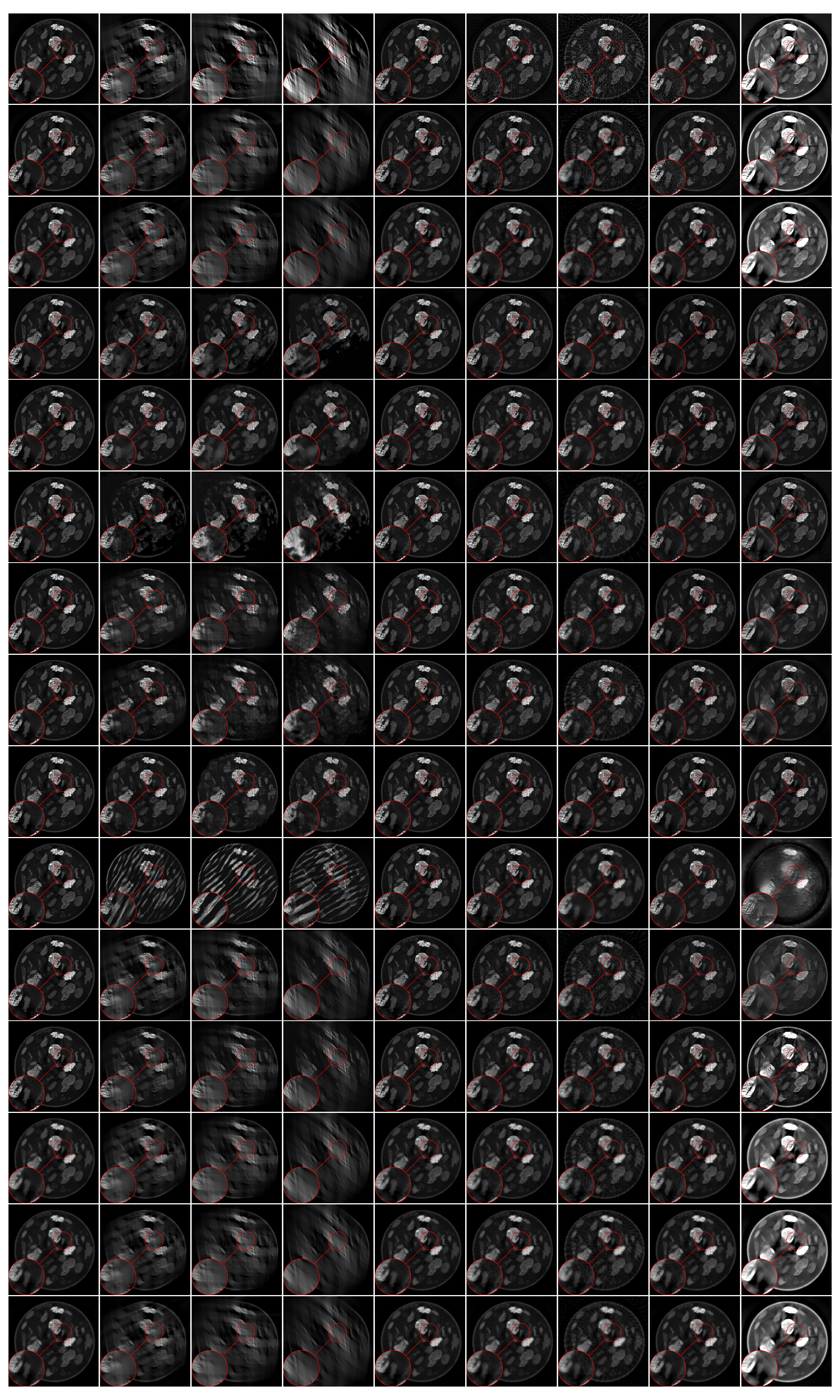}

\node[inner sep=0pt] (gt) at (-0.047,1.025)
    {\fcolorbox{green}{green}{\includegraphics[width=0.073\textwidth]{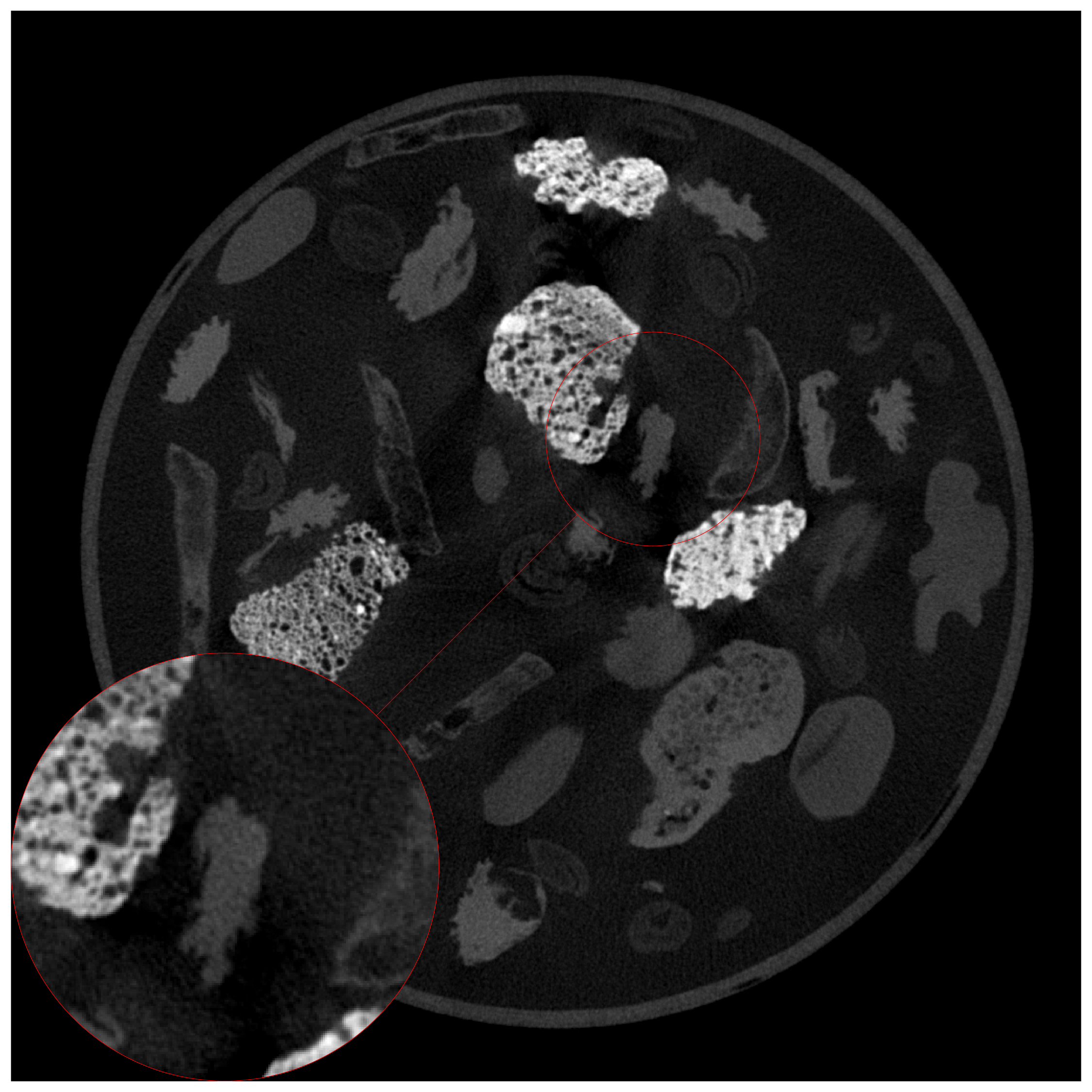}}};
% experiments
\node[rotate=70] at (0.06+0.01, 1.030) {\tiny{Full Data}};
\draw[gray] (0.1185,0.99) -- (0.1185,1.08);
\node[rotate=70]  at (0.17+0.01, 1.035) {\tiny{Limited 120}};
\node[rotate=70]  at (0.28+0.01, 1.032) {\tiny{Limited 90}};
\node[rotate=70]  at (0.39+0.01, 1.032) {\tiny{Limited 60}};
\draw[gray] (0.445,0.99) -- (0.445,1.08);

\node[rotate=70]  at (0.5+0.0, 1.032) {\tiny{Sparse 360}};
\node[rotate=70]  at (0.61+0.0, 1.032) {\tiny{Sparse 120}};
\node[rotate=70]  at (0.72+0.0, 1.030) {\tiny{Sparse 60}};
\draw[gray] (0.772,0.99) -- (0.772,1.08);

\node[rotate=70]  at (0.83+0.0, 1.032) {\tiny{Low-Dose}};
\node[rotate=70]  at (0.94+0.00, 1.05) {\tiny{Beam-Hardening}};

% algoritms

\node at (-0.02, 0.955) {\tiny{FBP}};
\node at (-0.0225, 0.955-0.065) {\tiny{AGD}};
\node at (-0.02, 0.955-0.065*2) {\tiny{ChP}};
\draw[gray] (-0.2,0.955-0.065*2.5+0.002) -- (0.01,0.955-0.065*2.5+0.002);

\node at (-0.065, 0.955-0.065*3) {\tiny{FBP+U-Net}};
\node at (-0.075, 0.955-0.065*4) {\tiny{FBP+MSDNet}};
\node at (-0.075, 0.955-0.065*5) {\tiny{FBP+DnCNN}};
\draw[gray] (-0.2,0.955-0.065*5.5+0.0005) -- (0.01,0.955-0.065*5.5+0.0005);

\node at (-0.015, 0.955-0.065*6) {\tiny{LG}};
\node at (-0.0275, 0.955-0.065*7) {\tiny{LGTV}};
\node at (-0.02, 0.955-0.065*8) {\tiny{LPD}};
\draw[gray] (-0.2,0.955-0.065*8.5-0.0005) -- (0.01,0.955-0.065*8.5-0.0005);

\node at (-0.015, 0.955-0.065*9) {\tiny{AR}};
\node at (-0.02, 0.955-0.065*10) {\tiny{TDV}};
\node at (-0.02, 0.955-0.065*11) {\tiny{ACR}};
\draw[gray] (-0.2,0.955-0.065*11.5-0.002) -- (0.01,0.955-0.065*11.5-0.002);

\node at (-0.07, 0.955-0.065*12) {\tiny{DnCNN-PnP}};
\node at (-0.075, 0.955-0.065*13) {\tiny{DRUNet-PnP}};
\node at (-0.05, 0.955-0.065*14) {\tiny{GS-PnP}};

\end{tikzonimage}
\caption{Qualitative analysis of all evaluated methods for slice 182 of the test dataset in comparison to the ``gold standard'' iterative reference reconstruction of the 2DeteCT dataset (green box).}
\label{fig:qualitative_analysis}
\end{figure}

We report on the performances of different data-driven methods on the most common CT image reconstruction tasks in both a quantitative (see Tables \ref{table:quantitative_analysis_modes}, \ref{table:quantitative_analysis_limited}, and \ref{table:quantitative_analysis_sparse}) and qualitative analysis (see Figure \ref{fig:qualitative_analysis}). The metrics are averaged over the whole test dataset and include their standard deviation whereas the qualitative analysis is for one specific slice (index 182) of the test dataset. While an extensive quantitative analysis of the performance describing trends in the different performances is out of the scope of this paper, we provide a detailed but short quantitative and qualitative analysis below. In CT image reconstruction the qualitative analysis, i.e. the visual inspection of reconstructed images, is a crucial tool to augment the quantitative results reported in Tables \ref{table:quantitative_analysis_modes}, \ref{table:quantitative_analysis_limited}, and \ref{table:quantitative_analysis_sparse} and shall serve as a starting point for further analysis.

\subsection*{Relevance and Difficulty of CT Image Reconstruction Tasks} \label{cha:disussion_relevance}
The basis of the benchmarking framework of this work are the selected CT image reconstruction tasks. Namely, Full Data, Limited Angle, Sparse Angle, Low-Dose and Beam-Hardening corrected reconstruction. In the following, we want to give a better insight about their respective relevance and difficulty.
\\[6pt]
\underline{Full Data Reconstruction:} The Full Data CT image reconstruction task can be considered purely as a reference for each of the algorithms and does not pose any particular challenges. This is due to the fact that this task uses the full data of the ``mode 2'' acquisition of the 2DeteCT dataset which was designed in a high-resolution setting with an over-sampling in the number of angular projections, a high-dose tube configuration, and with a beam filtration in place. Classical methods such as FBP, AGD, and ChP will perform well in these settings and there is no need for learned reconstruction methods.
\\[6pt]
\underline{Limited-Angle Reconstruction:} However, if this data is limited or sparsified in the angular range, undersampling artifacts occur. Theoretically, acquisitions from $180^\circ$ with sufficient angular sampling can produce an artifact-free image. When limiting the angular range further the missing information causes more visible image artifacts such as streaking, elongation, ghost tail, and missing boundaries. The challenge of limited angle acquisition occurs for example in industrial product inspection and medical imaging mammography. During the selection of the angular span of the wedge, we tested a limited angle of $150^\circ$ for a few algorithms and decided that the task at hand is not yet challenging enough. Therefore, we chose wedges of $120^\circ$, $90^\circ$, or $60^\circ$ for our limited angle reconstruction tasks. Since classical reconstructions of $60^\circ$ are already dominated by artifacts, a further undersampling was omitted. The difficulty of these tasks increases with decreasing available angular range.
\\[6pt]
\underline{Sparse-Angle Reconstruction:} For undersampling in terms of sparsity, the Nyquist-Shannon sampling theorem \cite{joseph1980view} can give an approximation of how many projections are necessary for a well-sampled CT scan. For the experimental setup of the 2DeteCT dataset a minimal number of approximately $3,000$ projections is required for sufficient sampling. Noticeable differences, however, only occur when undersampling by factors of five or more. Decreasing the number of projections is often used to speed up the CT acquisition process or to reduce dose for the scanned subject or sample. We tested specifically undersamplings based on $720$, $360$, $180$, $120$, $90$, and $60$ projections for one classical, one post-processing, and one unrolled algorithm to decide which experiments to include in the benchmark. Since both $720$ and $360$ projections showed a relatively similar severity in artifacts we chose to only include one of them in our benchmarking. For the lower end of this undersampling range we concluded that $60$ projections, so an undersampling of a factor 60 in comparison to the full data, was still feasible for some of the tested learned algorithms and should be considered as a challenging task for our benchmarking. To distribute the number of projections for our sparse angle reconstruction we chose $360$, $120$, and $60$ projections for our final experiments. Again, the difficulty of these tasks increases with decreasing the number of available projections.
\\[6pt]
\underline{Low-Dose Reconstruction:} For the low-dose CT image reconstruction task, we use the acquisition data of ``mode 1'' which uses a 1/30 tube current compared to ``mode 2'' acquisition that has been optimized for the best image quality. In medical imaging, a lower dose is typically chosen to achieve images of adequate quality for clinical purposes while minimizing radiation exposure. The ``tube current"-``exposure time'' products range from 50 to 400 mAs in clinical practice. ``Mode 2'' has a high-dose acquisition with a ``tube current''-``exposure time'' product of ~18 mAs, while ``mode 1'' has a low-dose acquisition with a product of 0.6 mAs. It is important to note that the size of the scanning object and the setup geometry differ from a traditional medical CT scan, as the scanning object is much smaller with a circumference of approximately 35 cm compared to 50-100 cm for standard abdominal circumferences in children and adolescents \cite{Reid2010}. Nevertheless, the low-dose CT reconstruction task can be viewed as having a similar or even higher noise-level as medical (extreme) low-dose CT data.
\\[6pt]
\underline{Beam-Hardening corrected Reconstruction:} For the beam-hardening corrected CT image reconstruction task, we use the acquisition data of the ``mode 3''. Since the beam spectrum of the X-ray source remains unfiltered during the acquisition process the corresponding iterative reconstruction slices show streaking and shadowing-like ``cupping'' artifacts as manifestations of beam hardening and photon starvation \cite{patton2008spect,van2002energy}. These artifacts are of a highly non-local and non-linear nature and corresponding data of beam-hardening afflicted CT images and physically filtered and corrected acquisition data are a novelty introduced by the 2DeteCT dataset \cite{Kiss_23}. In both medical and industrial settings having high-attenuating areas in the region-of-interest causes severe artifacts and poses challenges for real-world applications. The severity of the challenge of learning to map between these data distributions was to date unknown and is reported in this work for the first time. 
\\[6pt]
Overall, the limited-angle reconstruction from $60^\circ$ and the beam-hardening corrected reconstruction can be considered the most difficult tasks.

\begin{table}
  \caption{Quantitative analysis of all evaluated methods with respect to PSNR and SSIM on their performance in the CT image reconstruction tasks: Full Data ``mode 2'', Low-Dose ``mode 1'', and Beam-Hardening ``mode 3''.}
  \label{table:quantitative_analysis_modes}
  \centering
  \vspace{0.5cm}
  \resizebox{0.65\columnwidth}{!}{
  \begin{tabular}{ll ccc}
    \toprule
    \multirow{2}{*}{\textbf{Method}} & \multirow{2}{*}{\textbf{Metric}} & \multicolumn{3}{c}{\textbf{CT Image Reconstruction Task}}                   \\
    %\cmidrule(r){3-10} \\
     &  & Full Data & Low-Dose & Beam-Hardening \\
    \midrule
    \midrule
    
    \multicolumn{5}{l}{\textbf{Classical Methods}} \\
    \midrule
    \multirow{2}{*}{FBP} & SSIM      & 0.7463 $\pm$ 0.0296       & 0.0838 $\pm$ 0.0212      & 0.3367 $\pm$ 0.0464 \\
    & PSNR      & 35.0285 $\pm$ 2.0907       & 18.6437 $\pm$ 2.0508      & 14.5594 $\pm$ 1.9056 \\
    \midrule
    \multirow{2}{*}{AGD} & SSIM      & 0.7753 $\pm$ 0.0380      & 0.0727 $\pm$ 0.0182      & 0.3483 $\pm$ 0.0650 \\
    & PSNR      & \textbf{35.1006 $\pm$ 2.2801}      & 17.7062 $\pm$ 2.0517      & 14.4294 $\pm$ 1.9255 \\
    \midrule
    \multirow{2}{*}{ChP} & SSIM      & 0.7689 $\pm$ 0.0498     & 0.6820 $\pm$ 0.0749      & 0.4492 $\pm$ 0.0616 \\
    & PSNR      & 34.3251 $\pm$ 1.9703       & 31.9637 $\pm$ 1.9824      & 15.0090 $\pm$ 1.9094 \\
    \midrule
    \midrule
    
    \multicolumn{5}{l}{\textbf{Post-Processing Methods}} \\
    \midrule
    \multirow{2}{*}{FBP+U-Net} & SSIM      & 0.6499 $\pm$ 0.0681   & 0.7632 $\pm$ 0.0780      & 0.6336 $\pm$ 0.0760 \\
    & PSNR      & 32.6998 $\pm$ 1.9512       & 27.6772 $\pm$ 3.0133      & 28.0629 $\pm$ 3.8439 \\
    \midrule
    \multirow{2}{*}{FBP+MSDNet} & SSIM      & 0.8481 $\pm$ 0.0384   & 0.7253 $\pm$ 0.0864      & 0.7991 $\pm$ 0.0665 \\
    & PSNR      & 33.2999 $\pm$ 1.9492       & 31.6910 $\pm$ 1.9559      & 31.4389 $\pm$ 2.0988 \\
    \midrule
    \multirow{2}{*}{FBP+DnCNN} & SSIM      & 0.8324 $\pm$ 0.0403    & 0.7127 $\pm$ 0.0806      & 0.6328 $\pm$ 0.0751 \\
    & PSNR      & 32.2575 $\pm$ 1.9506      & 29.7645 $\pm$ 1.9748      & 28.2405 $\pm$ 2.0374 \\
    \midrule
    \midrule
    
    \multicolumn{5}{l}{\textbf{Learned / Unrolled Iterative Methods}} \\
    \midrule
    \multirow{2}{*}{LG} & SSIM      & 0.7498 $\pm$ 0.0708       & 0.6685 $\pm$ 0.0772      & 0.6025 $\pm$ 0.0747 \\
    & PSNR      & 32.4409 $\pm$ 1.9527       & 30.4679 $\pm$ 1.9945      & 28.4148 $\pm$ 1.9584 \\
    \midrule
    \multirow{2}{*}{LGTV} & SSIM      & 0.8221 $\pm$ 0.0532    & 0.7008 $\pm$ 0.0797      & 0.6656 $\pm$ 0.0774 \\
    & PSNR      & 33.3312 $\pm$ 1.9493     & 30.9991 $\pm$ 1.9328      & 29.5372 $\pm$ 2.0008 \\
    \midrule
    \multirow{2}{*}{LPD} & SSIM & 0.8447 $\pm$ 0.0370    & \textbf{0.8282 $\pm$ 0.0519}      & \textbf{0.8352 $\pm$ 0.0568} \\
    & PSNR & 33.3086 $\pm$ 1.9466  & \textbf{32.6685 $\pm$ 1.9656}      & \textbf{33.1382 $\pm$ 1.9759} \\
    \midrule
    \midrule

    \multicolumn{5}{l}{\textbf{Learned Regularizer Methods}} \\
    \midrule
    \multirow{2}{*}{AR} & SSIM & 0.8196 $\pm$ 0.0385  & 0.8039 $\pm$ 0.0505 & 0.3125 $\pm$ 0.0479 \\
    & PSNR & 32.3583 $\pm$ 1.9621 &31.0472 $\pm$ 1.9763 & 19.5692 $\pm$ 1.9612 \\
    \midrule
    \multirow{2}{*}{TDV} & SSIM & 0.7282 $\pm$ 0.0652 &  0.6047 $\pm$ 0.0815 & 0.5494 $\pm$ 0.0635\\
    & PSNR & 33.1204 $\pm$ 1.9496 &28.2429 $\pm$ 1.9433 & 25.7832 $\pm$ 2.0109 \\
    \midrule
    \multirow{2}{*}{ACR} & SSIM & 0.8518 $\pm$ 0.0362 & 0.8163 $\pm$ 0.0522 & 0.6742 $\pm$ 0.0505 \\
    & PSNR & 33.7131 $\pm$ 1.9450 & 32.2621 $\pm$ 1.9689 & 18.5708 $\pm$ 1.7824\\
    \midrule
    \midrule

    \multicolumn{5}{l}{\textbf{Plug-and-Play Methods}} \\
    \midrule
    \multirow{2}{*}{DnCNN-PnP} & SSIM & \textbf{0.8585 $\pm$ 0.0402} & 0.7795 $\pm$ 0.0523 & 0.5989 $\pm$ 0.0375 \\
    & PSNR & 32.8506 $\pm$ 1.9306 & 31.3986 $\pm$ 1.9424 & 15.4667 $\pm$ 1.9107 \\
    \midrule
    \multirow{2}{*}{DRUNet-PnP} & SSIM & 0.8573 $\pm$ 0.0405 & 0.7984 $\pm$ 0.0517 & 0.5945 $\pm$ 0.0375\\
    & PSNR & 32.8935 $\pm$ 1.9327 & 31.5762 $\pm$ 1.9480 & 15.4543 $\pm$ 1.9102 \\
    \midrule
    \multirow{2}{*}{GS-PnP} & SSIM & 0.7856 $\pm$ 0.0590  & 0.7727 $\pm$ 0.0683 & 0.5131 $\pm$ 0.0562\\
    & PSNR & 32.5734 $\pm$ 1.9331  & 31.7931 $\pm$ 1.9444 & 15.3466 $\pm$ 1.9128 \\
    \bottomrule
  \end{tabular}
  }
\end{table}

\begin{table}
  \caption{Quantitative analysis of all evaluated methods with respect to PSNR and SSIM on their performance in the CT image reconstruction tasks of Limited Angle Reconstruction.}
  \label{table:quantitative_analysis_limited}
  \centering
  \vspace{0.5cm}
  \resizebox{0.65\columnwidth}{!}{
  \begin{tabular}{ll ccc}
    \toprule
    \multirow{2}{*}{\textbf{Method}} & \multirow{2}{*}{\textbf{Metric}} & \multicolumn{3}{c}{\textbf{CT Image Reconstruction Task}}                   \\
    %\cmidrule(r){3-10} \\
     & & Limited Angle 120 & Limited Angle 90 & Limited Angle 60 \\
    \midrule
    \midrule
    
    \multicolumn{5}{l}{\textbf{Classical Methods}} \\
    \midrule
    \multirow{2}{*}{FBP} & SSIM      & 0.3418 $\pm$ 0.0354      & 0.2369 $\pm$ 0.0323      & 0.1557 $\pm$ 0.0288 \\
    & PSNR      & 22.4188 $\pm$ 1.9240      & 19.4251 $\pm$ 1.9405      & 16.9057 $\pm$ 1.9579     \\
    \midrule
    \multirow{2}{*}{AGD} & SSIM      & 0.4904 $\pm$ 0.0550      & 0.4411 $\pm$ 0.0555      & 0.4146 $\pm$ 0.0559 \\
    & PSNR     & 25.7508 $\pm$ 1.9690      & 24.1128 $\pm$ 1.9528      & 22.8848 $\pm$ 1.9531 \\
    \midrule
    \multirow{2}{*}{ChP} & SSIM      & 0.5923 $\pm$ 0.0550      & 0.5194 $\pm$ 0.0547      & 0.4658 $\pm$ 0.0510 \\
    & PSNR      & 26.3646 $\pm$ 1.9443      & 24.4392 $\pm$ 1.9321      & 22.7556 $\pm$ 1.9428 \\
    \midrule
    \midrule
    
    \multicolumn{5}{l}{\textbf{Post-Processing Methods}} \\
    \midrule
    \multirow{2}{*}{FBP+U-Net} & SSIM      & 0.7251 $\pm$ 0.0519      & 0.6338 $\pm$ 0.0659      & 0.5892 $\pm$ 0.0678 \\
    & PSNR      & 28.7931 $\pm$ 2.0052      & 27.3875 $\pm$ 2.0441      & 23.8511 $\pm$ 2.8370 \\
    \midrule
    \multirow{2}{*}{FBP+MSDNet} & SSIM      & 0.7840 $\pm$ 0.0641      & 0.7695 $\pm$ 0.0579      & 0.7148 $\pm$ 0.0726 \\
    & PSNR      & 30.7850 $\pm$ 1.9784      & 29.0111 $\pm$ 1.9740      & 27.4624 $\pm$ 1.9661 \\
    \midrule
    \multirow{2}{*}{FBP+DnCNN} & SSIM      & 0.5829 $\pm$ 0.0521      & 0.5661 $\pm$ 0.0528      & 0.5174 $\pm$ 0.0559 \\
    & PSNR      & 20.4433 $\pm$ 2.5345      & 23.4066 $\pm$ 2.2764      & 21.2132 $\pm$ 2.5140 \\
    \midrule
    \midrule
    
    \multicolumn{5}{l}{\textbf{Learned / Unrolled Iterative Methods}} \\
    \midrule
    \multirow{2}{*}{LG} & SSIM      & 0.6740 $\pm$ 0.0652      & 0.5746 $\pm$ 0.0655      & 0.5378 $\pm$ 0.0706 \\
    & PSNR      & 28.1639 $\pm$ 1.9380      & 26.3188 $\pm$ 1.9508      & 24.7228 $\pm$ 1.9630 \\
    \midrule
    \multirow{2}{*}{LGTV} & SSIM      & 0.6804 $\pm$ 0.0647      & 0.5590 $\pm$ 0.0639      & 0.5091 $\pm$ 0.0643 \\
    & PSNR      & 28.4131 $\pm$ 1.9300      & 26.0867 $\pm$ 1.9545      & 25.0225 $\pm$ 1.9687   \\
    \midrule
    \multirow{2}{*}{LPD} & SSIM & \textbf{0.8296 $\pm$ 0.0410}      & \textbf{0.8049 $\pm$ 0.0444}      & \textbf{0.7724 $\pm$ 0.0571} \\
    & PSNR & \textbf{31.1723 $\pm$ 1.9607}      & \textbf{29.2534 $\pm$ 1.9941}      & \textbf{28.0734 $\pm$ 1.9589} \\
    \midrule
    \midrule

    \multicolumn{5}{l}{\textbf{Learned Regularizer Methods}} \\
    \midrule
    \multirow{2}{*}{AR} & SSIM &0.6869 $\pm$ 0.0505 & 0.6100 $\pm$ 0.0543 &0.5742 $\pm$ 0.0620 \\
    & PSNR & 23.8496 $\pm$ 2.1578 & 21.1830 $\pm$ 2.1772 & 22.2350 $\pm$ 2.0260  \\
    \midrule
    \multirow{2}{*}{TDV} & SSIM & 0.5940 $\pm$ 0.0595 &0.5459 $\pm$ 0.0572 & 0.5282 $\pm$ 0.0584 \\
    & PSNR & 26.3233 $\pm$ 1.9315 &24.8127 $\pm$ 1.9399 & 23.3939 $\pm$ 1.9662 \\
    \midrule
    \multirow{2}{*}{ACR} & SSIM & 0.7114 $\pm$ 0.0543 & 0.6575 $\pm$ 0.0541& 0.5515 $\pm$  0.0529 \\
    & PSNR & 27.1792 $\pm$ 1.9441 & 25.3342 $\pm$ 1.9442 & 23.4915 $\pm$ 1.9539 \\
    \midrule
    \midrule

    \multicolumn{5}{l}{\textbf{Plug-and-Play Methods}} \\
    \midrule
    \multirow{2}{*}{DnCNN-PnP} & SSIM  & 0.7617 $\pm$ 0.0410  & 0.6981 $\pm$ 0.0441 &0.6200 $\pm$ 0.0475  \\
    & PSNR & 26.9997 $\pm$ 1.9330 & 25.0658 $\pm$ 1.9248 & 23.4108 $\pm$ 1.9521 \\
    \midrule
    \multirow{2}{*}{DRUNet-PnP} & SSIM & 0.7634 $\pm$ 0.0411 & 0.7002 $\pm$ 0.0443 & 0.6149 $\pm$ 0.0511 \\
    & PSNR  & 27.0262 $\pm$ 1.9334 & 25.0829 $\pm$ 1.9254 & 23.4362 $\pm$ 1.9520  \\
    \midrule
    \multirow{2}{*}{GS-PnP} & SSIM & 0.6396 $\pm$ 0.0670 & 0.5668 $\pm$ 0.0663 & 0.4989 $\pm$ 0.0625 \\
    & PSNR & 26.3318 $\pm$ 1.9328 & 24.5129 $\pm$ 1.9282 & 22.8225 $\pm$ 1.9481 \\
    \bottomrule
  \end{tabular}
  }
\end{table}

\begin{table}
  \caption{Quantitative analysis of all evaluated methods with respect to PSNR and SSIM on their performance in the CT image reconstruction tasks of Sparse Angle Reconstruction.}
  \label{table:quantitative_analysis_sparse}
  \centering
  \vspace{0.5cm}
  \resizebox{0.65\columnwidth}{!}{
  \begin{tabular}{ll ccc}
    \toprule
    \multirow{2}{*}{\textbf{Method}} & \multirow{2}{*}{\textbf{Metric}} & \multicolumn{3}{c}{\textbf{CT Image Reconstruction Task}}                   \\
    %\cmidrule(r){3-10} \\
     &  & Sparse Angle 360 & Sparse Angle 120 & Sparse Angle 60 \\
    \midrule
    \midrule
    
    \multicolumn{5}{l}{\textbf{Classical Methods}} \\
    \midrule
    \multirow{2}{*}{FBP} & SSIM      & 0.2947 $\pm$ 0.0453      & 0.1231 $\pm$ 0.0225      & 0.0611 $\pm$ 0.0112 \\
    & PSNR       & 24.9674 $\pm$ 2.0415      & 19.8769 $\pm$ 2.0124      & 16.6451 $\pm$ 1.9972 \\
    \midrule
    \multirow{2}{*}{AGD} & SSIM       & 0.3867 $\pm$ 0.0563      & 0.4142 $\pm$ 0.0630      & 0.4333 $\pm$ 0.0664 \\
    & PSNR      & 26.9629 $\pm$ 2.0444      & 27.5127 $\pm$ 1.9948      & 27.2796 $\pm$ 1.9553\\
    \midrule
    \multirow{2}{*}{ChP} & SSIM      & 0.6998 $\pm$ 0.0685      & 0.6712 $\pm$ 0.0718      & 0.5952 $\pm$ 0.0728  \\
    & PSNR      & 32.9158 $\pm$ 1.9739      & 31.7980 $\pm$ 1.9798      & 29.8020 $\pm$ 1.9326  \\
    \midrule
    \midrule
    
    \multicolumn{5}{l}{\textbf{Post-Processing Methods}} \\
    \midrule
    \multirow{2}{*}{FBP+U-Net} & SSIM      & 0.7449 $\pm$ 0.0801      & 0.7518 $\pm$ 0.0657      & 0.7728 $\pm$ 0.0592\\
    & PSNR      & 30.4766 $\pm$ 3.4844      & 26.0785 $\pm$ 6.3779      & 19.5421 $\pm$ 9.9613 \\
    \midrule
    \multirow{2}{*}{FBP+MSDNet} & SSIM      & 0.8392 $\pm$ 0.0473      & 0.7993 $\pm$ 0.0650      & 0.7626 $\pm$ 0.0789 \\
    & PSNR      & 33.1188 $\pm$ 1.9820      & 32.2993 $\pm$ 1.9928      & 30.9931 $\pm$ 1.9875      \\
    \midrule
    \multirow{2}{*}{FBP+DnCNN} & SSIM      & 0.7864 $\pm$ 0.0618      & 0.6701 $\pm$ 0.0864      & 0.6180 $\pm$ 0.0796       \\
    & PSNR      & 31.7575 $\pm$ 2.0213      & 29.1817 $\pm$ 2.3053      & 28.6079 $\pm$ 2.3389 \\
    \midrule
    \midrule
    
    \multicolumn{5}{l}{\textbf{Learned / Unrolled Iterative Methods}} \\
    \midrule
    \multirow{2}{*}{LG} & SSIM      & 0.7846 $\pm$ 0.0628      & 0.6795 $\pm$ 0.0790      & 0.6428 $\pm$ 0.0777 \\
    & PSNR      & 32.5946 $\pm$ 1.9764      & 31.1950 $\pm$ 1.9658      & 29.9360 $\pm$ 1.9603  \\
    \midrule
    \multirow{2}{*}{LGTV} & SSIM      & 0.7811 $\pm$ 0.0659      & 0.7100 $\pm$ 0.0745      & 0.7081 $\pm$ 0.0689 \\
    & PSNR      & 32.9404 $\pm$ 1.9666      & 31.4072 $\pm$ 1.9647      & 29.9021 $\pm$ 1.9357  \\
    \midrule
    \multirow{2}{*}{LPD} & SSIM & \textbf{0.8433 $\pm$ 0.0479}      & \textbf{0.8300 $\pm$ 0.0500}      & \textbf{0.8206 $\pm$ 0.0508}  \\
    & PSNR & \textbf{33.3809 $\pm$ 1.9513}      & \textbf{32.7032 $\pm$ 1.9685}      & \textbf{32.0583 $\pm$ 1.9789} \\
    \midrule
    \midrule

    \multicolumn{5}{l}{\textbf{Learned Regularizer Methods}} \\
    \midrule
    \multirow{2}{*}{AR} & SSIM & 0.8309 $\pm$ 0.0447  & 0.8117 $\pm$ 0.0553 &0.7949 $\pm$ 0.0595 \\
    & PSNR & 32.8067 $\pm$ 1.9714 & 32.1030 $\pm$ 1.9577 & 30.8378 $\pm$ 1.9532  \\
    \midrule
    \multirow{2}{*}{TDV} & SSIM & 0.6815 $\pm$ 0.0736 & 0.6235 $\pm$ 0.0741 &0.5725 $\pm$ 0.0728\\
    & PSNR & 32.2673 $\pm$ 1.9641 & 30.6585 $\pm$ 1.9357  & 28.9451 $\pm$ 1.8995\\
    \midrule
    \multirow{2}{*}{ACR} & SSIM & 0.8271 $\pm$ 0.0494 & 0.8074 $\pm$ 0.0539 &  0.7849 $\pm$  0.0524 \\
    & PSNR & 33.1537 $\pm$  1.9632 & 31.9181 $\pm$ 1.9666 & 30.5147$\pm$1.9338\\
    \midrule
    \midrule

    \multicolumn{5}{l}{\textbf{Plug-and-Play Methods}} \\
    \midrule
    \multirow{2}{*}{DnCNN-PnP} & SSIM  & 0.8405 $\pm$ 0.0432 & 0.8021 $\pm$ 0.0465 & 0.7637 $\pm$ 0.0484 \\
    & PSNR & 32.4627 $\pm$ 1.9309 & 31.3847 $\pm$ 1.9271 & 29.9350 $\pm$ 1.8980\\
    \midrule
    \multirow{2}{*}{DRUNet-PnP} & SSIM & 0.8398 $\pm$ 0.0433& 0.8000 $\pm$ 0.0465& 0.7658 $\pm$ 0.0498 \\
    & PSNR & 32.5065 $\pm$ 1.9324 & 31.3949 $\pm$ 1.9266 & 29.9518 $\pm$ 1.9085 \\
    \midrule
    \multirow{2}{*}{GS-PnP} & SSIM & 0.7622 $\pm$ 0.0628 & 0.7588 $\pm$ 0.0684 & 0.6937 $\pm$ 0.0701\\
    & PSNR & 32.1977 $\pm$ 1.9353 & 31.2728 $\pm$ 1.9370 & 29.5791 $\pm$ 1.8960  \\
    \bottomrule
  \end{tabular}
  }
\end{table}

\subsection*{Quantitative and Qualitative Analysis of the Evaluated Methods}
For this benchmarking we compared a range of algorithms representative for different categories of learned reconstruction methods in several CT image reconstruction tasks and reported their performance with respect to SSIM and PSNR. However, it is particularly important to note that the quantitative analysis presented in Tables \ref{table:quantitative_analysis_modes}, \ref{table:quantitative_analysis_limited}, and \ref{table:quantitative_analysis_sparse} does not capture the nuances of the quality of the reconstruction upon visual inspection. This is particularly true in the cases where the task at hand is more challenging, e.g. limited-angle reconstruction from $60^\circ$. In this case, the quantitative analysis indicates that both post-processing methods and learned/unrolled iterative methods perform similarly to the learned regularizer and PnP methods (e.g. ACR has better SSIM than LG). However, the qualitative analysis (visual inspection of the reconstructions) shows that for PnP and Learned Regularizer Methods, limited angle reconstructions are not performing well, arguably producing images as bad as FBP. Thus, one should not fully trust the performance metrics when comparing models solving such challenging inverse problem scenarios.
\\[6pt]
In the same line, the performance of all evaluated methods on the Full Data reconstruction task should be considered carefully. The ``gold standard'' or ground truth for this reconstruction task are iterative reference reconstructions computed with an AGD algorithm on a larger field-of-view ($2048\times2048$) and cropped to its center region of ($1024\times1024$). Given that this is a relatively well-posed problem, AGD should have converged to the true minimizer of the data fidelity functional, and thus the resulting image should be a good target for data-driven methods to have. However, this is not strictly true, as noise in CT acquisition is much more complex than just Gaussian, and this target is not really the exact ground truth. One could argue that a solution from an explicitly regularized algorithm (e.g. Chambolle-Pock with TV) would be also an appropriate target to use, and this would change the numerical results of this work. 
This choice of using AGD as target, is likely not highly impacting the conclusions of this work, but it is important to clarify that this is a choice, and not a definition of the ground truth.
\\[6pt]
This explains why AGD performs excellently, and if the same scale of the reconstruction would be kept, the SSIM would be 1 and PSNR infinity. But, as explained, this means that all evaluated methods in the Full Data reconstruction are learning to produce AGD-like results, not the actual ground truth. PnP and learned regularizer methods produce a high SSIM because they are also optimized using Gradient Descent, but with a learned regularization step. Therefore, they can mimic AGD more appropriately. 
\\[6pt]
The beam-hardening corrected reconstruction is a particularly interesting case study as the errors caused by beam-hardening are very non-local and non-linear. This presents a greater challenge than tasks which are more similar to ``denoising'' such as Sparse Angle. Therefore all methods that do not learn to imitate the final results directly (classical methods, PnP and adversarial regularizers) fail to produce a good image. Beam-hardening corrected reconstruction appears to be a more challenging task for variational regularization methods and learned regularizers and PnP methods. This can be explained by the linearization of the forward operator which assumes monochromatic X-ray sources. However, beam-hardening is ultimately a non-linear effect caused by wide beam spectra and their non-linear absorption. This operator mismatch pushes the minimization into the wrong direction, causing artifacts, as the forward model deviates from the physical process of acquiring the measurements. 
\\[6pt]
As an overall discussion on performance, it is worth noting that post-processing methods, albeit lacking mathematical guarantees, consistently produce quantitatively and visually relatively good results among all CT image reconstruction tasks. As is seen in Figure \ref{fig:CTReconstructionTasks}, for example for FBP+DnCNN in the Limited Angle reconstruction from $60^\circ$, these methods may suffer from ``hallucinations'', though. This should not come as a surprise as post-processing methods do not enforce data consistency. 
Generally, learned/unrolled iterative methods tend to be better at ensuring consistency in both data and image space, while (learned) regularization approaches provably achieve consistency \cite{scherzer2009variational}.
However, we emphasize that data consistency is necessary but not sufficient for preventing hallucinations, especially as the ill-posedness of the reconstruction problem increases, and as a result these methods may also suffer from hallucinations. Indeed, even classical, model-based, approaches with theoretical guarantees of data consistency have been seen to exhibit a sensitivity to adversarial perturbations \cite{genzel2022solving}. At the same time, training iterative unrolled methods requires much more time than post-processing methods, although the number of parameters are orders of magnitude smaller than standard networks like the U-Net.
\\[6pt]
Furthermore, we find that adversarial regularization and PnP approaches both tend to be well-performing in both sparse and full-data settings, oftentimes reaching performance of supervised learned/unrolled iterative methods, by the virtue of relying on the variational formulation and ultimately interpolating the missing image data. However, in the limited angle setting it is no longer an interpolation problem, but instead an in-painting problem. Neither PnP nor AR have been designed for in-painting directly and often rely on local image information. However, in-painting inherently requires non-local information in order to fill-in the missing data.

\subsection*{Limitations and Broader Impact} \label{sec:limitations}
In our benchmarking study, we aimed to establish a foundational understanding of how learned algorithms of different method categories perform on standardized CT reconstruction tasks with real-world experimental data. We prioritized adequate performance over extensive hyperparameter tuning to produce baseline results for a comparative analysis among techniques. However, we recognize that the interplay of hyperparameters — such as architecture, learning rates, regularization strengths, and iteration counts — can significantly affect performance. Consequently, the results shown in Figure \ref{fig:qualitative_analysis} and Tables \ref{table:quantitative_analysis_modes}, \ref{table:quantitative_analysis_limited}, and \ref{table:quantitative_analysis_sparse} should be interpreted cautiously, as they reflect performance under limited tuning. Future users must conduct thorough hyperparameter optimization tailored to their specific applications to fully leverage each method's potential.
\\[6pt]
While other CT image reconstruction tasks such as region-of-interest tomography, super-resolution, or segmentation are supported by the 2DeteCT dataset in principle, they have not yet been fully implemented in the benchmarking framework. Furthermore, while the dataset was designed to resemble abdominal CT scans, there are several remaining differences. However, 2DeteCT does provide realistic experimental data for a range of research fields such as manufacturing industry, food industry, and materials science. In an ideal case scenario, it would be possible to have raw measurement data for medical CT scanners and medically relevant subjects. However, medical CT manufacturers claim this data as proprietary and ethical concerns on both patients’ privacy and radiation dose prohibit acquiring matching data pairs of e.g. high-dose and low-dose scans or other acquisition modes. Moreover, the mismatch between the 2DeteCT dataset and medical CT image characteristics and morphology could create a significant performance gap that remains unexamined both in this study and in the existing literature. As a result, there is no guarantee that the overall performance trends observed in this work are transferable to medical CT cases. To address this, future research will need to focus on acquiring medical datasets and investigating retraining or transfer learning techniques to confirm or challenge these findings.
\\[6pt]
Additionally, the dataset has been acquired using a specific acquisition geometry and a non-medical micro-CT scanner, which could limit the generalization of trained algorithms to other CT data. For that, the performance of models trained on 2DeteCT could be evaluated under out-of-distribution (OOD) conditions in both image distribution and forward operator. For instance, applying the trained model to a different dataset would allow for assessing the models' generalization capacity. This type of experiment is particularly critical in the medical field, where out-of-distribution changes are common \cite{khorashadizadeh2024glimpse}.
\\[6pt]
Such an evaluation under OOD conditions should be conducted thoroughly, involving a wide range of OOD cases rather than just one or a few specific instances. While the extensive combinations of OOD tests and generalization scenarios poses significant computational challenges, making them impractical for the current study, we release the trained models and accompanying code on GitHub (see Section ``Code and Data Availability''). We hope this will facilitate future research in this area and believe that our comparison framework lays a solid foundation for conducting comprehensive OOD studies.
\\[6pt]
There are also slight remaining beam hardening artifacts in the filtered, clean acquisition of ``mode 2'', indicating a reduction but not complete removal of beam hardening. The chosen performance metrics of SSIM and PSNR are common quality assessments, but meaningful quality metrics for reconstructed (medical) CT images should be clinically relevant, task dependent, and aware of unaltered image content \cite{zhang2020review}. Despite these challenges, trained models on this dataset could potentially be applied to other data through transfer learning, with potential benefits for the medical sector. Researchers must be aware of potential distribution shifts and validate their algorithms on suitable data for the intended application. However, unlike other existing datasets, the 2DeteCT dataset provides raw experimental measurement data and the presented benchmarks show how well existing algorithms work on real-world experimental data. With this, we take a step towards closing the gap between developed algorithms and real-world applications by utilizing real-world experimental data, instead of simulated data.

\subsection*{Code and Data Availability} \label{sec:code_and_data}
The 2DeteCT dataset which serves as the foundation of this benchmarking paper is hosted on zenodo \cite{kiss_maximilian_b_2023_8014758} and the \href{https://github.com/CambridgeCIA/LION/}{LION toolbox} is hosted on GitHub by the Cambridge Image Analysis group and maintained by Ander Biguri. This open-source toolbox allows for easy access to a variety of methods, tools, and resources. Additionally, new methods can be seamlessly added to the toolbox, with demos available to showcase how the code should be organized and set up within the LION framework. This holds also true for future extensions of the available CT image reconstruction tasks mentioned in the paragraph ``Limitations and broader impact''. By regularly updating and expanding the toolbox the benchmarking project remains relevant, efficient, and effective in advancing the field of ML-based CT image reconstruction. The models and the scripts to train and evaluate the models presented in this benchmark, will be made available in the GitHub repository above. 

\section*{Conclusion}
Our benchmarking study provides a comparison of a fixed set of data-driven CT reconstruction algorithms with real-world experimental data in a reproducible and reusable way. It provides a starting point to develop new methods significantly faster as time-consuming implementations of data loaders, reconstruction tasks, comparison methods and evaluation protocols do not have to be redone. The open-source toolbox allows for seamless addition of new state-of-the-art methods and for extensions towards other problems and different CT reconstruction tasks.

\section*{Acknowledgments}
We deeply appreciate the help of Emilien Valat in setting up the data loader for the 2DeteCT dataset and the assistance of Johannes Krauß with the visualizations for this paper.

\newpage 
\bibliography{Benchmarking}

\end{document}